\documentclass[preprint2]{aastex}

\slugcomment{ApJ, in press}

\shorttitle{Radio Pulsar Search of $\gamma$-ray Binaries}
\shortauthors{McSwain et al.}

\begin{document}


\title{A Radio Pulsar Search of the $\gamma$-ray Binaries \\
LS I +61 303 and LS 5039}


\author{M.\ Virginia McSwain}
\affil{Department of Physics, Lehigh University, 16 Memorial Drive E, Bethlehem, PA 18015}
\email{mcswain@lehigh.edu}

\author{Paul S.\ Ray}
\affil{Space Science Division, Naval Research Laboratory, Code 7655, 4555 Overlook Avenue SW, Washington, DC, 20375}
\email{paul.ray@nrl.navy.mil}

\author{Scott M.\ Ransom}
\affil{National Radio Astronomy Observatory, 520 Edgemont Road, Charlottesville, VA 22903}
\email{sransom@nrao.edu}

\author{Mallory S.\ E.\ Roberts}
\affil{Eureka Scientific, Inc., 2452 Delmer Street, Suite 100, Oakland, CA 94602-3017}
\email{malloryr@gmail.com}

\author{Sean M.\ Dougherty}
\affil{National Research Council of Canada, Herzberg Institute for Astrophysics, Dominion Radio Astrophysical Observatory, PO Box 248, Penticton, British Columbia V2A 6J9, Canada}
\email{sean.dougherty@nrc.ca}

\author{Guy G.\ Pooley}
\affil{Cavendish Laboratory, University of Cambridge, J J Thomson Avenue, Cambridge CB3 0HE, UK}
\email{guy@mrao.cam.ac.uk}



\begin{abstract}
LS I +61 303 and LS 5039 are exceptionally rare examples of HMXBs with MeV--TeV emission, making them two of only five known or proposed ``$\gamma$-ray binaries''.  There has been disagreement within the literature over whether these systems are microquasars, with stellar winds accreting onto a compact object to produce high energy emission and relativistic jets, or whether their emission properties might be better explained by a relativistic pulsar wind colliding with the stellar wind.  Here we present an attempt to detect radio pulsars in both systems with the Green Bank Telescope.  The upper limits of flux density are between 4.1--14.5 $\mu$Jy, and we discuss the null results of the search.  Our spherically symmetric model of the wind of LS 5039 demonstrates that any pulsar emission will be strongly absorbed by the dense wind unless there is an evacuated region formed by a relativistic colliding wind shock.  LS I +61 303 contains a rapidly rotating Be star whose wind is concentrated near the stellar equator.  As long as the pulsar is not eclipsed by the circumstellar disk or viewed through the densest wind regions, detecting pulsed emission may be possible during part of the orbit.  
\end{abstract}





\section{Introduction}

LS I +61 303 is a high mass X-ray binary (HMXB) that is also a confirmed source of very high energy $\gamma$-ray emission.  The system consists of an optical star with spectral type B0 Ve and an unknown compact companion in a highly eccentric, 26.5 day orbit \citep{aragona2009, grundstrom2007, casares2005a}.  While the system has a relatively low X-ray luminosity for a HMXB, LS I +61 303 is the 15th brightest $\gamma$-ray source included in the \textit{Fermi} LAT 1-year Point Source Catalogue \citep{abdo2010}.  The Be disk interacts with the compact companion, producing orbital phase modulated emission across the electromagnetic spectrum:  TeV  \citep{albert2006, albert2008}, GeV \citep{abdo2009a}, X-ray \citep{paredes1997, leahy2001, torres2010}, optical H$\alpha$ \citep{mcswain2010}, and radio \citep{gregory1978, taylor1982, ray1997}. 
\citet{taylor1982} defined the arbitrary reference for zero phase at JD 2,443,366.775 that remains the conventional definition for LS I +61 303.  The binary orbital period ($P = 26.4960 \pm 0.0028$ d) has been determined from the periodic radio outbursts that peak near $\phi (\rm TG) = 0.6-0.8$ \citep{gregory2002}, and this period is confirmed by optical spectroscopy \citep{aragona2009}.  Periastron occurs at $\phi (\rm TG) = 0.275$ \citep{aragona2009}.  

LS 5039 is a HMXB that consists of a massive, main sequence star (ON6.5~V((f)) spectral type; \citealt{mcswain2004}) and a compact companion in a tightly bound, eccentric orbit \citep{casares2005b, casares2010, aragona2009}.   Similar to LS I +61 303, LS~5039 exhibits $\gamma$-ray \citep{abdo2009b, aharonian2006} and X-ray \citep{bosch2005, bosch2007, takahashi2009} emission that is variable over the 3.9 day orbital period.  There is also evidence for sub-orbital modulation of the stellar radial velocities, possibly due to tidally excited nonradial pulsations of the O-type star \citep{casares2010}.  

The systems LS I +61 303 and LS 5039 recently became subjects of controversy regarding the origin of their high energy emission and the nature of their compact companions.  Many studies (e.g.\ \citealt{massi2004},  \citealt{paredes2002}, \citealt{bosch2005}, \citealt{pandey2007}) have assumed the ``microquasar'' model -- so named because they are observed with extended radio emission that resemble small-scale versions of the relativistic jets of extragalactic quasars.  \citet{yamaguchi2010} argue that the keV--TeV variability of LS 5039 is due to to the injection of electrons into the relativistic jet in a low inclination binary orbit (which requires a black hole given the observed mass function; \citealt{aragona2009, casares2010}).  \citet{romero2007} argue that LS I +61 303 is an accreting microquasar based upon Smoothed Particle Hydrodynamics models of the stellar wind and the variable accretion rate during the orbit.  However, there is no observed evidence of an accretion disk in either system.  Optical spectroscopy may reveal H$\alpha$ emission and enhanced continuum emission from an accretion disk, but neither are observed in LS 5039 \citep{mcswain2004}.  Emission from the Be star's circumstellar disk dominates the H$\alpha$ line profile in LS I +61 303.  While we identified additional components of the H$\alpha$ emission from a spiral density wave in the Be disk and from a second interaction region, the emission is not consistent with an accretion disk \citep{mcswain2010}.  \citet{nagae2006} also did not detect polarization from a second disk in LS I +61 303.  High resolution X-ray spectroscopy should reveal an accretion disk from its thermal emission or a reflection iron line, but neither was detected in LS 5039 \citep{bosch2007}.  No high resolution grating spectra have been published for LS I +61 303, although an iron line is not required to fit low resolution spectra \citep{harrison2000, chernyakova2006}. 

Because the absence of an accretion disk is inconsistent with the microquasar model, \citet{dubus2006} proposed that the extended radio emission and observed X-ray/$\gamma$-ray properties of LS I +61 303 and LS 5039 are better explained by an interaction shock between the stellar wind and a relativistic wind from a non-accreting pulsar.  \citet{dhawan2006} argue that VLBA images covering a full orbit of LS I +61 303 support the pulsar model rather than a precessing jet.  H$\alpha$ emission from LS I +61 303 also supports the wind shock model \citep{mcswain2010}.  Such a model for LS I +61 303 was originally proposed by \citet{maraschi1981}.  The relativistic colliding wind mechanism successfully explains emission from PSR B1259$-$63 \citep{cominsky1994}, the first radio pulsar discovered in a binary system with a main sequence companion (B2 Ve star; \citealt{johnston1992}).  

Detecting a pulsar in LS I +61 303 or LS 5039 would conclusively settle the debate over the high energy production mechanism.  The pulsar timing solution would provide critical inputs for modeling, such as the magnetic field strength and spindown rate of the pulsar.  It would also provide a tomographic probe of the circumstellar envelope, as demonstrated with PSR B1259$-$63 \citep{johnston2005}.  Finally, the pulsar timing would also improve the precision of the orbital parameters that have been measured via optical spectroscopy.  

Several studies have attempted to detect the putative pulsars, to no avail.  LS 5039 lies within the strip of the Galactic plane surveyed by the Parkes Multibeam Pulsar Survey, which performed a low frequency search at 1374 MHz and flux density sensitivity $S_{1374} = 0.14$ mJy \citep{manchester2001}.  
LS I +61 303 is too far north to be included in the Arecibo L-band Feed Array (ALFA) pulsar survey \citep{cordes2006}, but it was included in the Green Bank Northern Sky Survey \citep{sayer1997}.  
X-ray pulse searches in LS I +61 303 have been performed by \citet{taylor1996} using \textit{RXTE}, \citet{harrison2000} with \textit{RXTE} and \textit{ASCA}, and by \citet{rea2010} with the \textit{Chandra} ACIS-S camera, each with null results.  


To further investigate the possible presence of a pulsar in these systems, we performed a deep radio pulse search modeled after successful observations of the radio pulsar PSR B1259$-$63 near its periastron \citep{johnston2005}.  Our observations were timed to occur near apastron, when the neutron star passes through a minimum in the surrounding stellar wind density.  
These observations are described in \S 2.  We model the stellar wind densities and optical depths, and discuss the null results from our pulsar search, in \S 3.


\section{Observations}

\begin{figure}
\includegraphics[angle=180,scale=0.4]{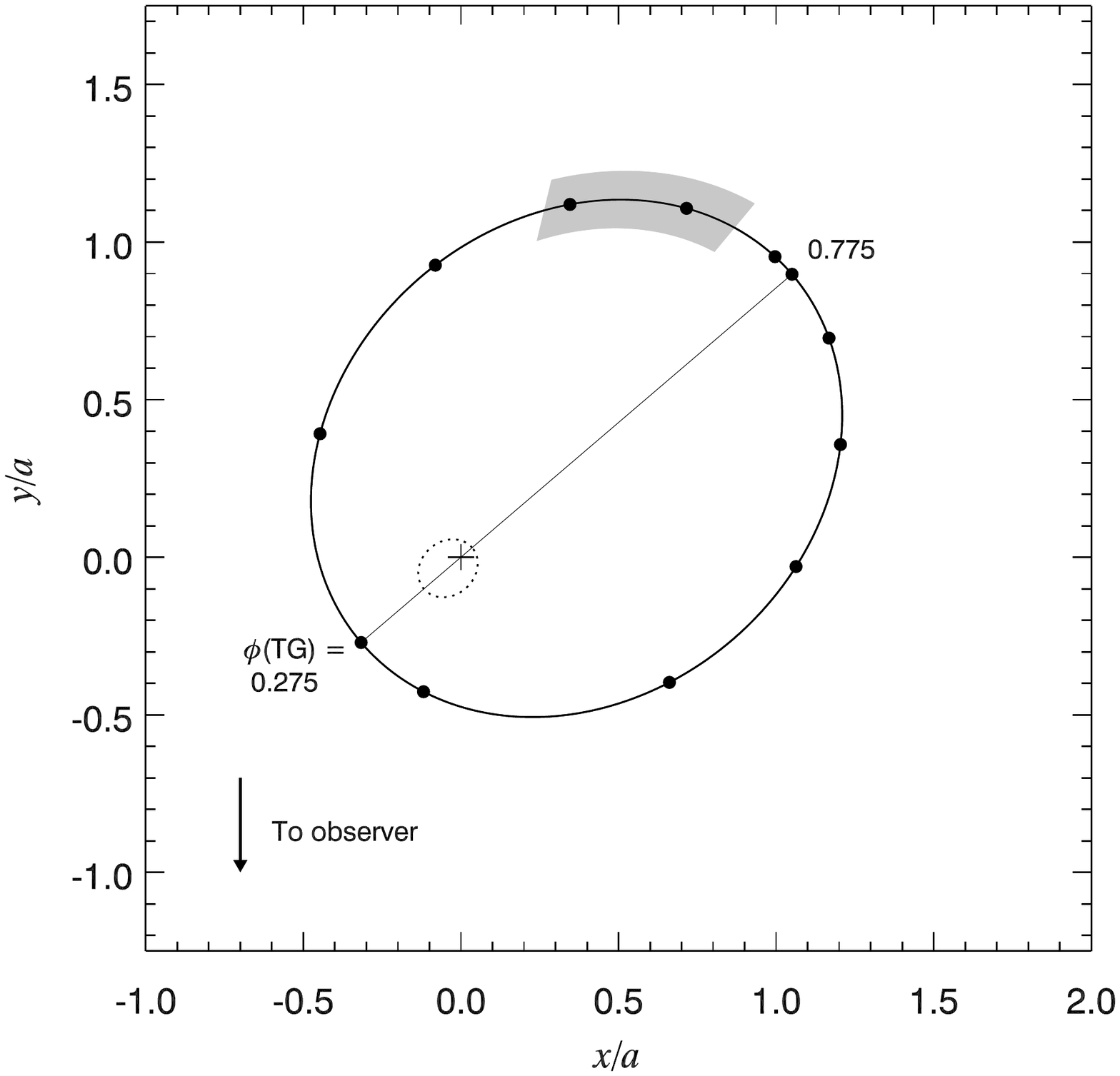} 
\caption{Orbital phase coverage of our GBT observations of LS I +61 303 (gray) compared to the orbital geometry \citep{aragona2009}.   The relative orbits ($r/a$) of the $12.5 \; M_\odot$ optical star (dotted line; \citealt{casares2005a}) and a presumed $1.4 \; M_\odot$ neutron star companion (solid line) are shown.  The center of mass is indicated with a cross, and the thin solid line is the orbital major axis.  Points around the orbit indicate steps of 0.1 in orbital phase, and the phases of periastron and apastron are labeled. 
\label{lsi_orbit} }
\end{figure}

\begin{figure}
\includegraphics[angle=180,scale=0.4]{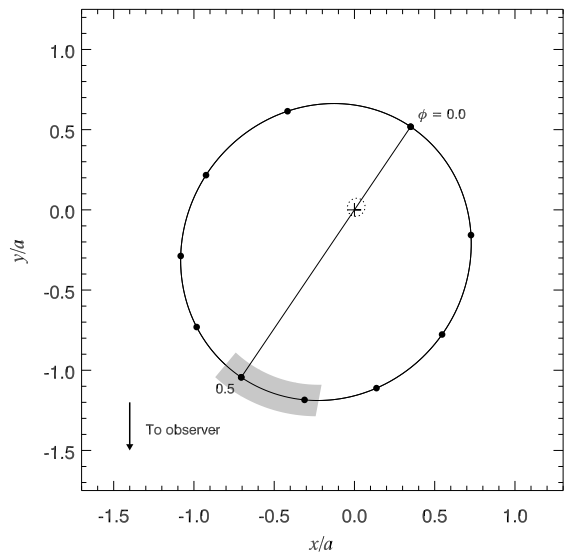} 
\caption{Orbital phase coverage of our GBT observations of LS~5039 (gray) compared to the orbital geometry, in the same format as Fig.\ \ref{lsi_orbit}.  Here, the optical star ($28.5 \; M_\odot$) orbits a presumed $1.4 \; M_\odot$ neutron star.  
\label{ls5039_orbit} }
\end{figure}

We observed LS I +61 303 and LS~5039 using the National Radio Astronomy Observatory's 100 m Green Bank Telescope (GBT) and the Pulsar Spigot back-end \citep{kaplan2005} in mode 2.  For LS I +61 303, we performed observations during 2008 May, September, and October using three receivers:  S-band with 600 MHz of bandwidth (1650--2250 MHz), C-band with 800 MHz of bandwidth (4400--5200 MHz), and X-band, also with 800 MHz of bandwidth (8500--9300 MHz).  We observed LS~5039 using only the lower frequency S-band (1700--2400 MHz) and C-band receivers during 2008 September and November.  Unfortunately we did not have sufficient time scheduled to also include the X-band for LS~5039.  The Spigot has 1024 available channels, but because of the smaller usable band at S-band, we only used 768 or 896 channels for those observations.  The Spigot summed and synthesized the data into 0.78125-MHz frequency channels every 81.92 $\mu$s.  The times of mid exposure and the corresponding orbital phases are listed in Table \ref{gbt}.  Illustrations of the orbital coverage of both systems are shown in Figures \ref{lsi_orbit} and \ref{ls5039_orbit}.  

\placetable{gbt}

\placefigure{lsi_orbit}

\placefigure{ls5039_orbit}

The data were reduced using the PRESTO software package \citep{ransom2001}.  We first searched the raw data for radio frequency interference in both the time and frequency domains and applied interference masks to remove any interference.  RFI was not a major issue with any of the searches except possibly for the S-band search of LS 5039, where approximately 20\% of the data were recommended to be removed.  For all the other observations the fraction was less than 2\% of the data.  As far as the fraction of the Fourier spectrum is concerned, we removed less than 0.02\% of the bins.  
LS I +61 303 and LS~5039 are expected to have interstellar dispersion measures (DM) of $\sim \!$ 100 cm$^{-3}$~pc using the NE2001 model \citep{cordes2002}.  We searched the observations by dedispersing the raw data into separate time series with DMs ranging from 0$-$2000 cm$^{-3}$~pc (to account for significant intersystem dispersion due to the dense stellar winds in both systems) and spaced by 0.5 cm$^{-3}$~pc.  We then Fourier-transformed each time series and searched them by using Fourier-domain acceleration search techniques in order to maintain sensitivity to the putative pulsars in their binary orbits.  For the acceleration search, we search a range of frequency derivatives, which are assumed to be constant over the observation.  This is an approximation to the true orbital acceleration of the binary system, which is sufficient for cases where the observation spans less than 10\% of the orbital period, as is the case here. Since it is possible that pulsed emission can only been detected in brief visibility windows and the DM may change significantly during the observations due to the clumped stellar winds, we also performed searches for single pulses using the standard tools in PRESTO.  We searched for pulse frequencies up to 1 kHz in each case.  

A typical spectral index of pulsars is $F_\nu \sim \nu^{-1.6}$ \citep{lorimer1995} or $\sim \nu^{-1.8}$ \citep{maron2000}, but the optical depth due to free-free absorption by the stellar winds decreases as $\tau_\nu \sim \nu^{-2}$ \citep{lamers1999}.  Therefore the high frequencies used in our search improve the sensitivity of our observations by countering the effects of radio absorption and scattering by the companions' winds.  
The flux density sensitivity limits are listed in Table \ref{gbt}, column 6.  
We did not detect any radio pulses from either LS I +61 303 or LS~5039.  


\section{Discussion}

For the purpose of this discussion, we assume that the compact companions in both LS I +61 303 and LS~5039 are $1.4 \; M_\odot$ neutron stars.  Using the mass functions, $f(m)$, and their formal errors determined by \citet{aragona2009}, we can constrain the inclination of both systems.  Assuming an optical companion with mass $M_\star = 12.5 \; M_\odot$ and radius $R_\star = 6.1 \; R_\odot$ in LS I +61 303 \citep{casares2005a}, the system inclination is $i = 73^\circ (+ 17^\circ, - 9^\circ)$.  
Likewise for LS~5039, we assume an optical companion of $28.5 \; M_\odot$ and $9.7 \; R_\odot$ based on the spectral type \citep{martins2005}, and $i = 71^\circ (+19^\circ, -8^\circ)$.  

The wind velocity, $v(r)$, of massive stars can be reasonably well estimated by the relation 
\begin{equation}
v(r) = v_\infty \left (1 - \frac{R_\star}{r} \right )^\beta
\end{equation}
\citep{lamers1999}.  For an O-type star with weak winds, as is LS 5039, the coefficient $\beta = 1$ \citep{puls1996}.    Stochastic clumping is ubiquitous among the winds of O stars, and we quantify these clumping effects using the method described by \citet{puls2008}.  The mean stellar wind density is 
\begin{equation}
\langle \rho (r) \rangle = \frac{\dot{M}}{ 4 \pi \, r^2 \, v(r)}.
\end{equation}
We assume that the space between clumps is a vacuum, and the density within the clumps, $\rho_c$, is enhanced by a volume filling factor $f_{\rm vol}$ such that 
\begin{equation}
\langle \rho \rangle = f_{\rm vol} \rho_c.
\end{equation}
The absorbing path length is reduced accordingly since no absorption occurs in the inter-clump region.  For LS 5039, we assume a spherically symmetric wind.  LS I +61 303 contains a rapidly rotating Be star, and below we discuss how the structure of the winds may be modified by the fast rotation.  

LS 5039 has a terminal wind velocity, $v_\infty$, of 2440~km~s$^{-1}$.  Its mass loss rate is $\dot{M} \approx 10^{-7} \; M_\odot$~yr$^{-1}$, although this rate is variable by a factor of 2 \citep{mcswain2004}.  LS 5039 is not known to be a rapidly rotating star, so no significant deviation from spherically symmetric winds is expected.  
Assuming a fully ionized hydrogen wind with a temperature of 2/3 the effective temperature of the O star ($T_{\rm eff} = 37500$ K; \citealt{mcswain2004}), the optical depth, $\tau_\nu \sim \nu^{-2}$, due to free-free absorption of the resulting wind density is given by \citet{lamers1999}.  Using this simple, spherically symmetric wind model for LS 5039, we integrated along the line of sight to the putative neutron star's position in the inclined orbit to determine the free-free optical depth as a function of orbital phase.  Since the wind is a highly ionized plasma, other scattering and/or absorption mechanisms may also be significant:  pulses may be smeared due to the variable column density or turbulence in the clumpy wind, nonlinear optical processes may scatter photons due to the high photon occupation number in the vicinity of a pulsar, and absorption at the cyclotron resonance frequency and its harmonics may be significant  \citep{thompson1994}.  We neglect these additional processes here.  

\begin{figure}
\hspace{-0.8in}
\includegraphics[angle=180,scale=0.4]{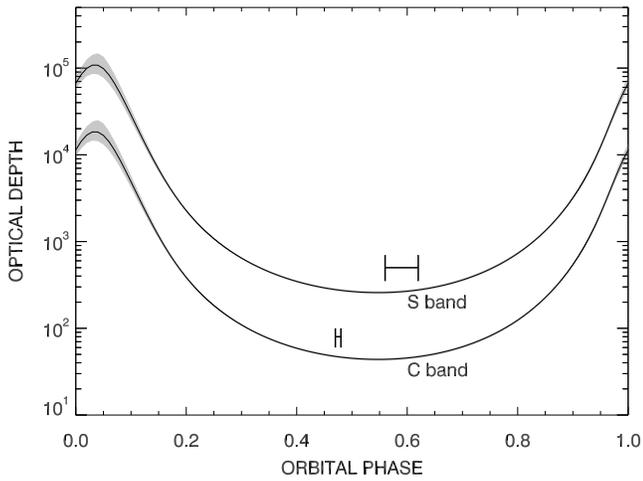} 
\caption{The calculated optical depth, $\tau_\nu$, of LS 5039 as a function of orbital phase.  Errors in $\tau_\nu$ due to the possible range in $i$ are shown in gray.  The orbital phases covered by our observations are marked for each receiver.  The calculation assumes a spherically symmetric wind but neglects the effects of an interacting wind shock region.  
\label{ls5039_optdepth} }
\end{figure}

Our resulting $\tau_\nu$ for LS 5039 are plotted in Figure \ref{ls5039_optdepth} for $f_{\rm vol} = 1$ (corresponding to an unclumped wind).  Since the free-free opacity has a $\rho^2$ dependence, the effect of wind clumping is to enhance $\tau_\nu$ by a factor of $1/f_{\rm vol}$ despite the reduced path length of absorbers.  Evolved, Galactic O stars have $0.1 < f_{\rm vol} < 0.5$ (\citealt{puls2008} and references therein), although the less evolved, main sequence O star in LS 5039 may have less strongly clumped winds.  Since LS 5039's mass loss rate is variable, the predicted optical depth may also vary by a factor of $\sim 2$ at the time of our observations. 
The large $\tau_\nu \gg 1$ are prohibitive for detecting any possible pulsar in the LS 5039 system at the S-band and C-band frequencies used in our search.  

\placefigure{ls5039_optdepth}

However, we have neglected the effects of a conical wind shock region, which likely shields the putative pulsar in LS 5039 from the dense stellar wind.  If the O star's winds are colliding with a relativistic pulsar wind, then the wind collision zone would result in a double shock structure separated by a contact discontinuity (CD; \citealt{bogovalov2008}).  The structure has a conical shape that envelopes the compact object or the hot star, depending on the ratio of momentum flux of the winds,  $\eta = (\dot{E}/c) / (\dot{M} v)$.  This ratio uniquely determines the opening angle and location of the shock structure, and it will define the topography of a wind cavity bounded by the CD.   At the time of inferior conjunction, the absorption along the neutron star's line of sight may be far less than predicted for a uniform stellar wind.  High resolution, high signal-to-noise, phase-resolved observations of the ultraviolet wind resonance lines are necessary to determine how large a cavity is evacuated within the stellar winds given the fast orbit of LS 5039. 

\begin{figure}
\hspace{-0.75in}
\includegraphics[angle=180,scale=0.4]{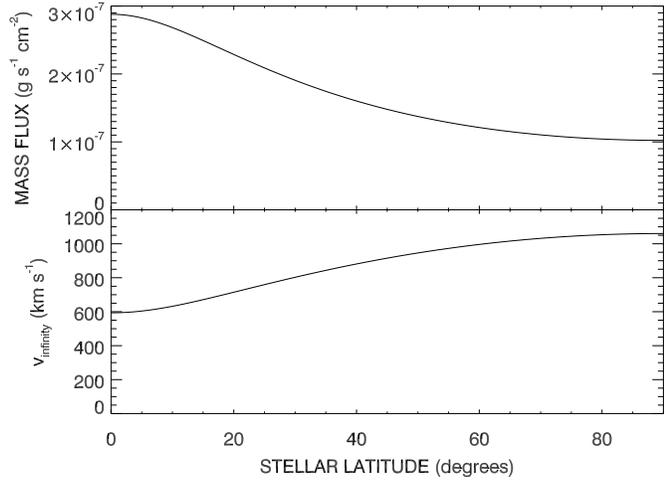} 
\caption{The mass flux (top) and terminal wind velocity $v_\infty$ (bottom) for LS I +61 303, assuming a rotationally distorted stellar wind.  The cooler $T_{\rm eq, eff}$ leads to decreased equatorial radiative flux but increased opacity, enhancing the equatorial mass loss rate.  The mass flux is normalized to a total $\dot{M} = 10^{-8} \; M_\odot$~yr$^{-1}$.  The lower radiation flux decreases the wind momentum and thus $v_\infty$ at the equator.  
\label{lsi_massloss} }
\end{figure}

The wind properties of LS I +61 303 are uncertain, but we estimated its $\dot{M} \approx 10^{-8} \; M_\odot$~yr$^{-1}$ using the method of \citet{lamers1999} for B-type stars.  The optical star in this system is a rapidly rotating Be star, probably rotating near the critical velocity at which gravity balances the centrifugal force at the equator.  For a typical Be star, $V_{\rm rot}/V_{\rm crit} \sim 0.8$ \citep{mcswain2008}.  The fast rotation leads to several interesting effects at the stellar surface that affect the mass loss.  Due to the large centrifugal force at the equator, the Roche model predicts that the equatorial radius $R_{\rm eq}$ is about 1.3 times larger than the polar radius $R_{\rm p}$ \citep{maeder2009}.  According to the von Zeipel theorem, the smaller effective surface gravity at the equator implies a cooler equatorial temperature, $T_{\rm eq, eff}$, and decreased equatorial flux \citep{vonzeipel1924}.  We modeled LS I +61 303 using a polar temperature of $T_{\rm p, eff} = 30,000$ K and $T_{\rm eq, eff} = 21,500$ K, with a mean surface temperature $T_{\rm eff} = 25,100$ K.  This $T_{\rm eff}$ is consistent with measurements by \citet{howarth1983} and references therein.  The cooler $T_{\rm eq, eff}$ leads to an increased opacity, enhancing the equatorial mass loss rate despite the lower radiative flux \citep{maeder2009}.  In Figure \ref{lsi_massloss}, we plot the resulting mass flux as a function of stellar latitude, normalized to a total $\dot{M} = 10^{-8} \; M_\odot$~yr$^{-1}$.  The wind momentum is also affected by the variation of radiative flux with latitude \citep{maeder2009}, leading to a slower wind velocity at the equator (also shown in Figure \ref{lsi_massloss}).  Combining these effects, we find a significantly denser wind at the equator, so our model is a reasonable representation of the dense circumstellar disk of the Be star.  Finally, we assume that the equatorial disk is aligned with the binary orbit, we neglect wind clumping effects, and we integrate through the resulting wind distribution along the line of sight to the putative pulsar to calculate $\tau_\nu$ during the orbit.  

\placefigure{lsi_massloss}

\begin{figure}
\hspace{-0.8in}
\includegraphics[angle=180,scale=0.4]{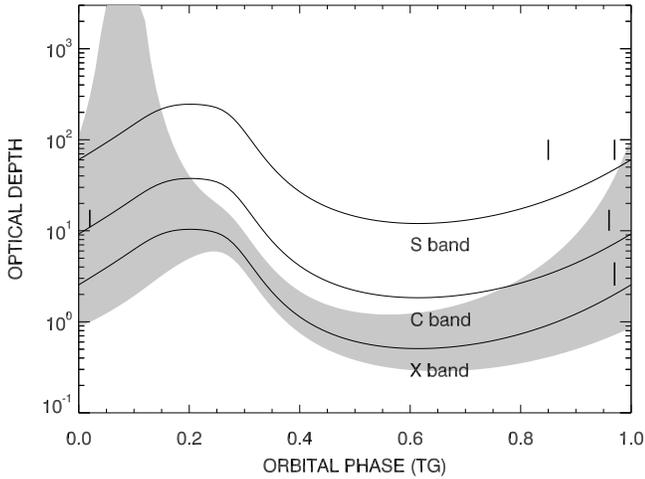} 
\caption{The calculated optical depth, $\tau_\nu$, of LS I +61 303 as a function of orbital phase, using the ephemeris of \citet{taylor1982}.  Errors in $\tau_\nu$ due to the possible range in $i$ are shown in gray for the X band only; the C and S band errors are comparable.  The orbital phases covered by our observations are marked for each receiver.  The calculation assumes a rotationally distorted star whose equatorial plane is aligned with the orbital plane, and it neglects the effects of an interacting wind shock region. 
\label{lsi_optdepth} }
\end{figure}

We plot the resulting $\tau_\nu$ as a function of orbital phase for LS I +61 303's asymmetric wind in Figure \ref{lsi_optdepth}.  Only the X-band provides an opportunity for an optically-thin view of the pulsar during $0.4 \le \phi (\rm TG) \le 0.8$, assuming the system inclination is $i = 73^\circ$.  Due to telescope scheduling constraints, our GBT X-band observation was performed at $\phi (\rm TG) = 0.97$.  While our model suggests that $\tau_X < 1$ for the lower limit of $i$, it is not an optimal phase for detecting pulses.  If the inclination is high, the pulsar will be viewed through the densest portion of the wind and would likely be highly absorbed at all orbital phases.  

\placefigure{lsi_optdepth}

While the Be star's disk is truncated to a maximum radius of 34--37 $R_\odot$ \citep{grundstrom2007}, we predict that the pulsar can be eclipsed by the disk if the orbit is viewed edge-on, leading to very high $\tau_\nu$ near $\phi (\rm TG) = 0.1$.  However, in this post-supernova system, there is no requirement that the orbital plane remain aligned with the circumstellar disk.  A kick velocity from the supernova would likely have disrupted the system's angular momentum sufficiently to misalign the orbit.  Indeed, the Be star disk in PSR B1259$-$63 is highly inclined relative to the orbital plane \citep{johnston1999, bogomazov2005}.  Therefore a pulsar in LS I +61 303 may not be strongly eclipsed at any phase.  Further observations to constrain both the stellar inclination as well as the orbital inclination are necessary to improve our model of the stellar wind density of LS I +61 303.  


Based on our model, we expect that the free-free absorption is prohibitively large at the S-band and C-band frequencies to detect a pulsar at any orbital phase of either  LS I +61 303 or LS~5039.  The high frequency X-band offers the best chance of countering the effects of stellar wind absorption.  
Although we did not detect pulsars in this search, we conclude that detecting a pulsar in LS~5039, using higher frequencies, may be possible if the CD due to a colliding wind shock region effectively shields the pulsar from the dense stellar wind.  Detecting a pulsar in LS I +61 303 should be possible using the X-band during a more optimal orbital phase as long as the pulsar beam crosses our line of sight.  

In order to power these $\gamma$-ray sources, any associated pulsar should have a high spin-down power $\gtrsim 10^{36}$ erg~s$^{-1}$.  Such a pulsar likely has a rotation period $\sim$ 0.05--0.4 s, with an average beaming fraction of 20--50\% \citep{tauris1998}.  
There are only four known $\gamma$-ray binaries with well established MeV--TeV emission:  LS I +61 303, LS~5039, PSR B1259$-$63, and HESS J0632+057 \citep{aharonian2007, bongiorno2011}.  PSR B1259$-$63 remains the only system with a detected radio pulsar, but with further observations it may be possible to expand this elite group.  

Future X-ray pulsation searches of these sources are also encouraged.  We note that the HMXB 4U 0115+63 (= V635 Cas) has a very similar spectral type and orbital period compared to LS I +61 303 (B0.2 Ve, 24.3 d in this case; \citep{liu2006}) and their stellar wind properties should be very similar.  That system is known to include an accreting pulsar (e.g.\ \citealt{nakajima2010}).  The thermal emission of the hot star winds, usually with $0.2 \le kT \le 0.6$ keV but sometimes hotter \citep{naze2011} will likely prohibit the detection of thermal surface emission from a neutron star.  Harder, non-thermal X-rays (above 2 keV) should be more effective for pulsation searches in these $\gamma$-ray binaries.


\acknowledgments

We are grateful to Carl Bignell and the staff at NRAO for their help scheduling and performing these observations.  We also thank Guillaume Dubus and Anna Szostek for helpful discussions about the stellar winds and possible contact discontinuities in these systems.  The National Radio Astronomy Observatory is a facility of the National Science Foundation operated under cooperative agreement by Associated Universities, Inc.  This work is supported by NASA DPR numbers NNX08AV70G, NNG08E1671, and NNX09AT67G.  MVM is grateful for an institutional grant from Lehigh University.



{\it Facilities:} \facility{GBT}.


\clearpage

\clearpage

\begin{deluxetable}{lcccccc}
\tablewidth{0pt}
\tablecaption{Journal of GBT Observations \label{gbt} }
\tablehead{
\colhead{} &
\colhead{MJD of} &
\colhead{Orbital} &
\colhead{Exposure Time} &
\colhead{ } &
\colhead{$S_{\rm lim}$} \\
\colhead{Star} &
\colhead{Mid Exposure} &
\colhead{Phase}         &
\colhead{(hr)} &
\colhead{Receiver} & 
\colhead{($\mu$Jy)} }
\startdata
LS I +61 303	&  54599.631	&  0.96  		&  3.0  &  C-band  & \phn 4.1  \\  
			&  54599.777	&  0.97  		&  3.0  &  X-band  & \phn 7.5  \\  
			&  54599.882	&  0.97  		&  1.2  &  S-band  & \phn 9.4  \\  
			&  54729.006	&  0.85  		&  0.5  &  S-band  &        14.5  \\  
			&  54760.010	&  0.02  		&  1.5  &  C-band  & \phn 5.8  \\  
LS 5039		&  54730.051	&  0.56--0.62  	&  4.4  &  S-band  & \phn 5.2  \\  
			&  54796.010	&  0.47--0.48  	&  1.5  &  C-band  & \phn 5.9  \\  
\enddata
\end{deluxetable}

\end{document}